\documentclass[12pt]{article}

\usepackage{graphicx}
\usepackage{amsmath}

\def\ltwid{\mathrel{\raise.3ex\hbox{$<$\kern-.75em\lower1ex\hbox{$\sim$}}}}
\def\gtwid{\mathrel{\raise.3ex\hbox{$>$\kern-.75em\lower1ex\hbox{$\sim$}}}}

\def\square{\kern1pt\vbox{\hrule height 1.2pt\hbox{\vrule width 1.2pt\hskip 3pt
   \vbox{\vskip 6pt}\hskip 3pt\vrule width 0.6pt}\hrule height 0.6pt}\kern1pt}
\def\overleftrightarrow#1{\vbox{\ialign{##\crcr
     $\leftrightarrow$\crcr\noalign{\kern-1pt\nointerlineskip}
     $\hfil\displaystyle{#1}\hfil$\crcr}}}

\begin{document}

\begin{titlepage}

\begin{flushright}
UFIFT-QG-15-05
\end{flushright}

\vskip 1cm

\begin{center}
{\bf Fine Tuning May Not Be Enough}
\end{center}

\vskip 1cm

\begin{center}
S. P. Miao$^{1*}$ and R. P. Woodard$^{2\dagger}$
\end{center}

\begin{center}
\it{$^1$ Department of Physics, National Cheng Kung University \\ 
No.1, University Road, Tainan City 70101, TAIWAN}
\end{center}

\begin{center}
\it{$^{2}$ Department of Physics, University of Florida \\
Gainesville, FL 32611, UNITED STATES}
\end{center}

\vskip .5cm

\begin{center}
ABSTRACT
\end{center}

We argue that the fine tuning problems of scalar-driven inflation
may be worse than is commonly believed. The reason is that reheating
requires the inflaton to be coupled to other matter fields whose
vacuum fluctuations alter the inflaton potential. The usual response 
has been that even more fine-tuning of the classical potential 
$V(\varphi)$ can repair any damage done in this way. We point out 
that the effective potential in de Sitter background actually depends 
in a complicated way upon the dimensionless combination of $\varphi/H$.
We also show that the factors of $H$ which occur in de Sitter do not 
even correspond to local functionals of the metric for general 
geometries, nor are they Planck-suppressed.

\begin{flushleft}
\end{flushleft}

\vskip 1cm

\begin{flushleft}
$^*$ e-mail: spmiao5@mail.ncku.edu.tw \\
$^{\dagger}$ e-mail: woodard@phys.ufl.edu
\end{flushleft}

\end{titlepage}

\section{Introduction}

It is desirable to define inflation as a phase of accelerated
expansion, distinct from any particular mechanism for achieving
it. The explanatory power of an epoch accelerated expansion during 
the very early universe was realized in the late 1970's and early
1980's \cite{Brout:1977ix,Starobinsky:1980te,Kazanas:1980tx,
Sato:1980yn,Guth:1980zm}. The case for primordial inflation is 
now so strong as to be nearly incontrovertible \cite{Kinney:2012,
Woodard:2009ns}. However, there is no similarly compelling 
indication for what caused it to occur.

The earliest, and still the simplest, model of inflation is general 
relativity with a minimally coupled scalar inflaton $\varphi(x)$ 
\cite{Linde:1981mu,Albrecht:1982wi,Linde:1983gd},
\begin{equation}
\mathcal{L} = \frac{R \sqrt{-g}}{16 \pi G} -\frac12 \partial_{\mu}
\varphi \partial_{\nu} \varphi g^{\mu\nu} \sqrt{-g} - V(\varphi)
\sqrt{-g} \; . \label{singlescalar}
\end{equation}
Models of this type certainly work because there is constructive
procedure for determining the potential $V(\varphi)$ needed to
support any expansion history \cite{Tsamis:1997rk,Saini:1999ba,
Capozziello:2005mj,Woodard:2006nt,Guo:2006ab}. However, there are
six fine-tuning problems:
\begin{enumerate}
\item{{\it Initial Conditions} --- Inflation will not begin unless
the inflaton is approximately homogeneous, with more potential energy 
than kinetic energy, over more than a Hubble volume 
\cite{Goldwirth:1989pr,Goldwirth:1991rj,Vachaspati:1998dy};}
\item{{\it Duration} --- Inflaton will not last long enough (more 
than 50 e-foldings) unless $V(\varphi)$ is very flat 
\cite{Linde:1981mu,Albrecht:1982wi};}
\item{{\it Scalar Perturbations} --- The magnitude of scalar 
perturbations will not agree with observations unless $(G V)^3/{V'}^2 
\sim 10^{-11}$ \cite{Mukhanov:1990me};}
\item{{\it Tensor Perturbations} --- The magnitude of tensor
perturbations will be too large unless ${V'}^2/G V^2 \ltwid 1$ 
\cite{Mukhanov:1990me};}
\item{{\it Reheating} --- The post-inflationary universe will not
reach a hot enough temperature unless the inflaton couples to 
ordinary matter because its gravitational couplings do not suffice
\cite{Allahverdi:2010xz};} 
\item{{\it Cosmological Constant} --- The post-inflationary 
universe will not evolve correctly unless the minimum of the scalar 
potential obeys $G^2 V_{\rm min} \approx 10^{-123}$ 
\cite{Riess:1998cb,Perlmutter:1998np,Wang:2006ts,Alam:2006kj}.}
\end{enumerate}

Some of these conditions work against one another. In particular,
the matter couplings required by (5) induce Coleman-Weinberg 
corrections of the form $\pm \varphi^4 \ln(\varphi)$ which disturb
conditions (2-4) and necessitate additional fine tuning. The 
staggering amount of fine-tuning which is required disturbs many 
cosmologists \cite{Ijjas:2013vea,Guth:2013sya,Linde:2014nna,
Ijjas:2014nta}.

However distasteful all this fine tuning might seem, it has always
been believed that the thing could at least be done. The purpose of 
this paper is to point out that this may not be true: the matter-inflaton 
couplings which are required by (5) actually induce gravitational 
couplings which are not Planck-suppressed and are not even local. 
Because allowed modifications of the action are restricted to be local, 
these nonlocal corrections to the inflaton effective potential cannot 
be eliminated. In section 2 we describe the true form of the effective 
potentials which emerge on de Sitter background from other scalars, 
from Yukawa fermions and from gauge particles. In section 3 we argue 
that the complicated factors of the de Sitter Hubble constant $H$ which 
appear in these results are not local functionals of the metric for a
general cosmological geometry. Section 4 discusses the practical 
problem.

\section{Effective Potentials for de Sitter}

The purpose of this section is to give the one loop quantum corrections
to the inflaton potential arising from hypothetical couplings to other 
scalars, to a massless fermion, and (for a charged inflaton) to a vector
gauge boson. In each case we first give a formal expression for the 
result in terms of the unknown coincidence limit of the propagator for
massive scalars with arbitrary conformal coupling, 
\begin{equation}
\Bigl[ \square \!-\! \xi R \!-\! M^2\Bigr] i\Delta[\xi,M^2](x;x')
= \frac{i \delta^D(x \!-\! x')}{\sqrt{-g}} \; , \label{genprop}
\end{equation}
where $R$ is the Ricci scalar and $\square$ is the covariant scalar 
d'Alembertian $\sqrt{-g} \, \square \equiv \partial_{\mu} (\sqrt{-g} \, 
g^{\mu\nu} \partial_{\nu})$. The only inflationary geometry for which 
the coincidence limit of this propagator is known for $M \neq 0$ is de 
Sitter,
\begin{equation}
i\Delta[\xi,M^2](x;x) \Bigl\vert_{\rm de\ Sitter} = \frac{H^{D-2}}{(
4\pi)^{\frac{D}2}} \frac{\Gamma( \frac{D-1}2 \!+\! \nu) \Gamma(\frac{D-1}2
\!-\! \nu) \Gamma(1 \!-\! \frac{D}2)}{\Gamma(\frac12 \!+\! \nu) \Gamma(
\frac12 \!-\! \nu)} \; , \label{dSprop}
\end{equation}
where $\nu^2 \equiv (\frac{D-1}2)^2 - D (D-1) \xi -\frac{M^2}{H^2}$ and
$D$ is the dimension of spacetime, which is kept arbitrary to facilitate
dimensional regularization. In each case (scalar, fermion and vector) we 
give the full de Sitter result, along with the large field and small field 
expansions.

\subsection{Coupling to another scalar}

The inflaton might couple to another scalar $\phi(x)$, which need not be
minimally coupled to gravity,
\begin{equation}
\mathcal{L}_{\rm scalar} = -\frac12 \partial_{\mu} \phi \partial_{\nu} \phi
g^{\mu\nu} \sqrt{-g} - \frac1{12} \Bigl(1 \!+\! \Delta \xi\Bigr) \phi^2 
R \sqrt{-g} - \frac14 h^2 \phi^2 \varphi^2 \sqrt{-g} \; . \label{scalarL}
\end{equation}
At one loop order this coupling induces the following correction to the
derivative of the inflaton potential (assuming, as always, that $\varphi$ 
is constant),
\begin{equation}
\Delta V'_{\rm scalar}(\varphi) = \delta \xi R \varphi + \frac16 \delta 
\lambda \varphi^3 + \frac12 h^2 \varphi \, i\Delta[\xi,\frac12 h^2 
\varphi^2](x;x) \; .
\end{equation} 
The coincidence limit of the propagator is the primitive contribution,
while $\delta \xi R \varphi$ represents a renormalization of the 
inflaton's (classically zero) conformal coupling and $\frac16 \delta
\lambda \varphi^3$ renormalizes the inflaton's quartic self-coupling. 

We specialize to de Sitter and choose the finite parts of the two 
counterterms to cancel the $\varphi^2$ and $\varphi^4$ terms in the small 
field expansion \cite{Bilandzic:2007nb,Janssen:2009pb},
\begin{eqnarray}
\delta \xi & \!\!\!\!=\!\!\!\! & - \frac{\Delta \xi h^2 H^{D-4} 
\Gamma(1 \!-\! \frac{D}2)}{12 (4\pi)^{\frac{D}2}} - 
\frac{\Delta \xi h^2}{192 \pi^2} \Bigl[ \psi( \nu_+) + \psi( \nu_-) 
\Bigr] \; , \label{dxiscalar} \\
\delta \lambda & \!\!\!\!=\!\!\!\! & - \frac{3 h^4 H^{D-4} 
\Gamma(1 \!-\! \frac{D}2)}{2 (4\pi)^{\frac{D}2}} - 
\frac{3 \Delta \xi h^4}{16 \pi^2} \Bigl[ \frac{\psi( \nu_+) \!+\! 
\psi( \nu_-)}{2 \Delta \xi} - \frac{[ \psi'( \nu_+) \!-\! 
\psi'( \nu_-)]}{\sqrt{1 \!-\! 8 \Delta \xi}} \Bigr] \; . \quad
\label{dlambdascalar}
\end{eqnarray}
Here and henceforth $\psi(x) \equiv \frac{d}{dx} \ln[\Gamma(x)]$ is the
digamma function, and we define $\nu_{\pm} \equiv \frac12 \pm \frac12 
\sqrt{1 - 8 \Delta \xi}$. The renormalized result can be expressed in 
terms of the dimensionless quantity $z \equiv h \varphi/H$ 
\cite{Bilandzic:2007nb,Janssen:2009pb},
\begin{eqnarray}
\lefteqn{ \hspace{-.3cm} \Delta V_{\rm scalar} = \frac{H^4}{64 \pi^2} 
\Biggl\{ \!\!-\Bigl[ \psi(\nu_+\!) \!+\! \psi( \nu_-) \Bigr] \! \Bigl[2 
\Delta \xi z^2 \!\!+\! \frac{z^4}4\Bigr] \!+\! \Bigl[ \psi'( \nu_+\!) 
\!-\! \psi'( \nu_-) \Bigr] \frac{\frac12 \Delta \xi z^4}{\sqrt{1 \!-\! 8 
\Delta \xi}} } \nonumber \\
& & \hspace{-.5cm} + \!\! \int_0^{z^2} \!\!\!\! dx \Bigl(2 \Delta \xi \!+\! 
\frac{x}2 \Bigr) \Biggl[ \psi\Bigl( \frac12 \!+\! \sqrt{\frac14 \!-\! 
2 \Delta \xi \!-\! \frac{x}2} \, \Bigr) \!+\! \psi\Bigl( \frac12 \!-\! 
\sqrt{\frac14 \!-\! 2 \Delta \xi \!-\! \frac{x}2} \, \Bigr) \Biggr] 
\! \Biggr\} . \quad \label{scalarDV}
\end{eqnarray}

The large field expansion comes from substituting in (\ref{scalarDV}) the
large argument expansion for the digamma function,
\begin{equation}
\vert z\vert \gg 1 \Longrightarrow \psi(z) = \ln(z) - \frac1{2 z} - 
\frac1{12 z^2} + \frac1{120 z^4} - \frac1{256 z^6} + O\Bigl(\frac1{z^8}
\Bigr) \; . \qquad \label{largepsi}
\end{equation}
The resulting expansion is,
\begin{eqnarray}
\lefteqn{ \Delta V_{\rm scalar} = \frac{H^4}{64 \pi^2} \Biggl\{ \frac14
z^4 \ln\Bigl(\frac12 z^2 \!+\! 2 \Delta \xi\Bigr) - \Biggl[ \frac18 + 
\frac{[\psi(\nu_+) \!+\! \psi(\nu_-)]}{4} } \nonumber \\
& & \hspace{-.5cm} - \frac{\Delta \xi [\psi'(\nu_+) \!-\! \psi'(\nu_-)]}{
2 \sqrt{1 \!-\! 8 \Delta \xi}} \Biggr] z^4 + 2 \Delta \xi z^2 
\ln\Bigl( \frac12 z^2 \!+\! 2 \Delta \xi\Bigr) - \Biggl[\frac13 \!+\! 
\Delta \xi \nonumber \\
& & \hspace{0cm} + 2 \Delta \xi \Bigl[ \psi(\nu_+) \!+\! \psi(\nu_-)
\Bigr]\Biggr] z^2 + \Bigl[4 \Delta \xi^2 \!-\! \frac2{15}\Bigr] 
\ln\Bigl( \frac12 z^2 \!+\! 2 \Delta \xi\Bigr) + O(z^0) \Biggr\} . 
\qquad \label{largescalar}
\end{eqnarray}
Because taking $H$ to zero makes $z = h \varphi/H$ large, it is this form
(\ref{largescalar}) which makes contact with the famous Coleman-Weinberg 
potential of flat space \cite{Coleman:1973jx}. To get the small field
expansion we Taylor expand the digamma functions in expression 
(\ref{scalarDV}) and integrate,
\begin{eqnarray}
\lefteqn{ \Delta V_{\rm scalar} = \frac{H^4}{64 \pi^2} \Biggl\{ \Biggl[
\frac{(1 \!-\! 6 \Delta \xi) [-\psi'(\nu_+) \!+\! \psi'(\nu_-)]}{(1 \!-\!
8 \Delta \xi)^{\frac32}} + \frac{ \Delta \xi [\psi''(\nu_+) \!+\! 
\psi''(\nu_-)]}{1 \!-\! 8 \Delta \xi} \Biggr] \frac{z^6}{12} } 
\nonumber \\
& & \hspace{-.3cm} + \Biggl[\frac{3 (1 \!-\! 4 \Delta \xi) [-\psi'(\nu_+) 
\!+\! \psi'(\nu_-)]}{(1 \!-\! 8 \Delta \xi)^{\frac52}} +
\frac{3 (1 \!-\! 4 \Delta \xi) [\psi''(\nu_+) \!+\! \psi''(\nu_-)]}{2 (1 
\!-\! 8 \Delta \xi)^{2}} \nonumber \\
& & \hspace{4.7cm} + \frac{\Delta \xi [-\psi'''(\nu_+) \!+\! \psi'''(\nu_-)]
}{(1 \!-\! 8 \Delta \xi)^{\frac32}} \Biggr] \frac{z^8}{96} + O(z^{10}) 
\Biggr\} . \qquad
\end{eqnarray}

\subsection{Coupling to a massless fermion}

The inflaton might be Yukawa-coupled to a massless Dirac fermion $\psi_i(x)$
\begin{equation}
\mathcal{L}_{\rm fermion} = \overline{\psi} \gamma^b e^{\mu}_{~ b} 
\Bigl( \partial_{\mu} \!+\! \frac{i}2 A_{\mu cd} J^{cd} \Bigr) \psi
\sqrt{-g} - f \varphi \overline{\psi} \psi \sqrt{-g} \; .
\label{fermionL}
\end{equation}
Here $e^{\mu}_{~b}(x)$ is the vierbein field with $g^{\mu\nu}(x)
= e^{\mu}_{~b}(x) e^{\nu}_{~c}(x) \eta^{bc}$ and $A_{\mu cd}(x) = 
e^{\nu}_{~c} [e_{\nu d , \mu} - \Gamma^{\rho}_{~\mu\nu} e_{\rho d}]$
is the spin connection. The symbol $\gamma^b_{ij}$ represents the $4 
\times 4$ gamma matrices which obey $\{ \gamma^b,\gamma^c\} = -2 
\eta^{bc} I$, and $J^{cd} \equiv \frac{i}{4} [\gamma^c,\gamma^d]$ 
are the Lorentz representation matrices for Dirac fermions. 

For cosmology we only require the fermion propagator for a general 
homogeneous, isotropic and spatially flat geometry in conformal 
coordinates,
\begin{equation}
ds^2 = a^2(\eta) \Bigl[-d\eta^2 + d\vec{x} \!\cdot\! d\vec{x}\Bigr]
\quad e_{\mu b}(x) = a(\eta) \eta_{\mu b} \quad H(\eta) \equiv
\frac{a'}{a^2} \quad \epsilon(\eta) \equiv -\frac{H'}{a H^2} \; ,
\label{cosmogeom}
\end{equation}
where $H(\eta)$ is the Hubble parameter and $\epsilon(\eta)$ is the
first slow roll parameter, which is assumed to lie in the range 
$0 \leq \epsilon < 1$. In the cosmological geometry (\ref{cosmogeom})
the appropriate fermion propagator is \cite{Candelas:1975du,Miao:2006pn},
\begin{eqnarray}
\lefteqn{i \Bigl[\mbox{}_i S_j\Bigr](x;x') = \frac1{a^{\frac{D+1}2}(\eta)}
\Bigl[ i \gamma^{\mu} \partial_{\mu} + a(\eta) m I\Bigr]
\frac{ a^{\frac{D-1}2}(\eta)}{\sqrt{a(\eta) a(\eta')} } } \nonumber \\
& & \hspace{1.3cm} \times \Biggl\{i
\Delta[\xi_c,M^2_+](x;x') \Bigl( \frac{I \!+\! \gamma^0}2\Bigr)
+ i \Delta[\xi_c,M^2_-](x;x') \Bigl( \frac{I \!-\! \gamma^0}2\Bigr)
\Biggr\} , \qquad 
\end{eqnarray}
where $\xi_c \equiv \frac1{2(D-1)}$ and $M^2_{\pm} \equiv f \varphi
(f \varphi \mp i H)$.

At one loop order the coupling (\ref{fermionL}) induces the following 
correction to the $\varphi$ derivative of the inflaton potential,
\begin{equation}
\Delta V'_{\rm fermion}(\varphi) = \delta \xi \varphi R + \frac16
\delta \lambda \varphi^3 - f i\Bigl[\mbox{}_i S_i\Bigr](x;x) \; .
\label{DVfermion}
\end{equation}
Of course the trace of the coincident fermion propagator is the 
primitive contribution, while the terms proportional to $\delta \xi$
and $\delta \lambda$ are counterterms. Specializing to de Sitter 
(hence $H(\eta)$ constant) and renormalizing so as to null the
quadratic and quartic terms in the small field expansion
\cite{Miao:2006pn},
\begin{eqnarray}
\delta \xi & = & \frac{4 f^2 H^{D-4}}{(4\pi)^{\frac{D}2}} \, 
\frac{\Gamma(1 \!-\! \frac{D}2)}{D (D \!-\! 1)} + 
\frac{(1 \!-\! \gamma) f^2}{24 \pi^2} \; , \label{dxifermion} \\
\delta \lambda & = & \frac{24 f^4 H^{D-4}}{(4 \pi)^{\frac{D}2}}
\, \Gamma\Bigl(1 \!-\! \frac{D}2\Bigr) + \frac{3 [\zeta(3) \!-\! 
\gamma] f^4}{\pi^2} \; , \label{dlambdafermion}
\end{eqnarray}
gives rise to the following renormalized result 
\cite{Candelas:1975du,Miao:2006pn},
\begin{eqnarray}
\Delta V_{\rm fermion}(\varphi) 
& = & -\frac{H^4}{8 \pi^2} \Biggl\{2 \gamma \Bigl(\frac{f\varphi}{H}\Bigr)^2
- [\zeta(3) \!-\! \gamma] \Bigl(\frac{f\varphi}{H}\Bigr)^4 \nonumber \\
& & \hspace{2cm} + 2 \int_0^{\frac{f\varphi}{H}} \!\!\!\!\! dx \, (x \!+\!
x^3) \Bigl[\psi(1 \!+\! i x) \!+\! \psi(1 \!-\! i x)\Bigr] \Biggr\} . \qquad
\label{fermionDV}
\end{eqnarray}

Substituting (\ref{largepsi}) in (\ref{fermionDV}) gives the large field
expansion with $z \equiv f \varphi/H$,
\begin{eqnarray}
\lefteqn{\Delta V_{\rm fermion} = -\frac{H^4}{8 \pi^2} \Biggl\{ 
\frac12 z^4 \ln(z^2 \!+\! 1) - \Bigl( \zeta(3) \!+\! \frac14 \!-\! \gamma\Bigr) 
z^4 + z^2 \ln(z^2 \!+\! 1)} \nonumber \\
& & \hspace{4.8cm} - \Bigl( \frac43 \!-\! 2\gamma\Bigr) z^2 + \frac{11}{60}
\ln(z^2 \!+\! 1) + O(z^0) \Biggr\} . \qquad \label{largefermion}
\end{eqnarray}
As before, the Coleman-Weinberg form \cite{Coleman:1973jx} is apparent in
the leading large $z$ behavior. To get the small field expansion we 
substitute,
\begin{equation}
\vert z \vert \ll 1 \Longrightarrow \psi(1 \!+\! z) = - \gamma -
\sum_{k = 1}^{\infty} \zeta(k \!+\! 1) (-z)^k \; . \qquad \label{smallpsi}
\end{equation}
The resulting expansion is,
\begin{eqnarray}
\Delta V_{\rm fermion} & = & -\frac{H^4}{4 \pi^2} \sum_{n=2}^{\infty} 
\frac{(-1)^n}{n \!+\! 1} \Bigl[ \zeta(2n \!-\! 1) - \zeta(2n \!+\! 1)\Bigr] 
z^{2n+2} \; , \qquad \\
& = & -\frac{H^4}{8\pi^2} \Biggl\{ \frac23 \Bigl[\zeta(3) \!-\! \zeta(5)\Bigr]
z^6 -\frac1{2} \Bigl[\zeta(5) \!-\! \zeta(7)\Bigr] z^8 + O(z^{10})\Biggr\} .
\qquad 
\end{eqnarray}

\subsection{Coupling to a vector gauge boson}

A complex inflaton might couple to electromagnetism,
\begin{equation}
\mathcal{L}_{\rm vector} = -\frac14 F_{\rho\sigma} F_{\mu\nu} g^{\rho \mu}
g^{\sigma \nu} \sqrt{-g} - \Bigl( \partial_{\mu} \!-\! ie A_{\mu}\Bigr) \varphi 
\Bigl(\partial_{\nu} \!+\! i e A_{\nu}\Bigr) \varphi^* g^{\mu\nu} \sqrt{-g} \; .
\label{vectorL}
\end{equation}
Here $A_{\mu}(x)$ is the vector potential and $F_{\mu\nu} \equiv \partial_{\mu}
A_{\nu} - \partial_{\nu} A_{\mu}$ is the field strength tensor. At our one loop 
level the result for a non-Abelian charged scalar would be proportional to the 
electromagnetic result derived from (\ref{vectorL}).

Coleman and Weinberg noted that Lorentz gauge ($\partial_{\mu} [\sqrt{-g} 
g^{\mu\nu} A_{\nu}] = 0$) makes the coupling $ie A_{\mu} [\varphi^*_{, \nu} 
\varphi - \varphi^* \varphi_{, \nu}] g^{\mu\nu} \sqrt{-g}$ drop from the one 
loop effective potential \cite{Coleman:1973jx}. Hence we employ the transverse 
vector propagator \cite{Tsamis:2006gj},
\begin{equation}
\Bigl[ \square^{\mu\nu} \!-\! R^{\mu\nu} \!-\! M_V^2 g^{\mu\nu} \Bigr] 
i\Bigl[\mbox{}_{\nu} \Delta_{\rho}\Bigr](x;x') = \frac{i \delta^{\mu}_{\rho} 
\delta^D(x \!-\! x')}{\sqrt{-g}} + g^{\mu\nu} \partial_{\nu} \partial'_{\rho} 
 i\Delta[0,0](x;x') \; ,
\end{equation}
where $\square^{\mu\nu}$ is the covariant vector d'Alembertian and $M_V^2 = 
2 e^2 \varphi^* \varphi$. For de Sitter (and we conjecture generally) it is 
best to express this propagator as 2nd order transverse projectors on 
$x^{\mu}$ and ${x'}^{\mu}$, contracted into an invariant bi-vector multiplied 
by a scalar structure function \cite{Miao:2011fc},
\begin{eqnarray}
i\Bigl[\mbox{}_{\mu} \Delta_{\rho}\Bigr](x;x') & = & 
\mathbf{P}_{\mu}^{~\nu}(x) \!\times\! \mathbf{P}_{\rho}^{~\sigma}(x') 
\!\times\! \Biggl[\frac{\partial^2 \ell^2(x;x')}{\partial x^{\nu} 
\partial {x'}^{\sigma}} \!\times\! \mathcal{S}(x;x') \Biggr] \; , \qquad \\
\mathbf{P}_{\mu}^{~\nu}(x) & \equiv & \square_{\mu}^{~\nu} - D^{\nu}
D_{\mu} \; , \\
\mathcal{S}(x;x') & \equiv & \frac{i \Delta[\xi_v,M_V^2] \!-\!
i\Delta[\xi_v,0]}{M_V^4} -\frac1{M^2_V} \frac{\partial i\Delta[\xi_v,N^2]}{
\partial N^2} \Biggr\vert_{N^2 = 0} \; . \qquad
\end{eqnarray}
Here $D_{\mu}$ is the covariant derivative operator, $\xi_v \equiv 
\frac{(D-2)}{D (D-1)}$ and $\ell^2(x;x')$ is some function of the invariant 
length from $x^{\mu}$ to ${x'}^{\mu}$.

The photon loop contribution to the inflaton potential takes the form,
\begin{equation}
\Delta V'_{\rm vector}(\varphi^* \varphi) = \delta \xi R + \frac12 \delta \lambda
\varphi^* \varphi + e^2 g^{\mu\nu} \, i\Bigl[\mbox{}_{\mu} \Delta_{\nu}\Bigr](x;x)
\; . \label{DVvector}
\end{equation}
As before, we choose the conformal and quartic renormalizations to cancel the
$\varphi^* \varphi$ and $(\varphi^* \varphi)^2$ terms in the small field
expansion of $V_{\rm vector}(\varphi^* \varphi)$ \cite{Prokopec:2007ak},
\begin{eqnarray}
\delta \xi & = & \frac{e^2 H^{D-4}}{(4\pi)^{\frac{D}2}} \Biggl\{ \frac1{4\!-\!D}
+ \frac{\gamma}2 + O(D \!-\!4) \Biggr\} \; , \label{dxivector} \\
\delta \lambda & = & \frac{D (D\!-\!1) e^4 H^{D-4}}{(4\pi)^{\frac{D}2}} \Biggl\{
\frac2{4 \!-\! D} + \gamma - \frac32 + O(D \!-\! 4)\Biggr\} \; . 
\label{dlambdavector}
\end{eqnarray}
The final renormalized result is \cite{Prokopec:2007ak,Allen:1983dg},
\begin{eqnarray}
\lefteqn{\Delta V_{\rm vector} = \frac{3 H^4}{8 \pi^2} \Biggl\{ \Bigl(-1 \!+\! 
2 \gamma\Bigr) z^2 + \Bigl(-\frac32 + \gamma\Bigr ) z^4 } \nonumber \\
& & \hspace{1.5cm} + \int_0^{z^2} \!\!\! dx \, (1 \!+\! x) \Biggl[ 
\psi\Bigl(\frac32 \!+\! \frac12 \sqrt{1 \!-\! 8 x}\Bigr) + 
\psi\Bigl(\frac32 \!-\! \frac12 \sqrt{1 \!-\! 8x}\Bigr)\Biggr] \Biggr\} , 
\qquad \label{vectorDV}
\end{eqnarray}
where we define $z^2 \equiv \frac{e^2 \varphi^* \varphi}{H^2}$.

The large field expansion derives from substituting (\ref{largepsi}) in
(\ref{vectorDV}),
\begin{eqnarray}
\lefteqn{\Delta V_{\rm vector} = \frac{3 H^4}{8 \pi^2} \Biggl\{ \frac12 z^4 
\ln(z^2 \!+\! 1) + \Bigl[-\frac74 + \frac12 \ln(2) + \gamma\Bigr] z^4 + z^2 
\ln(z^2 \!+\! 1)} \nonumber \\
& & \hspace{3cm} + \Bigl[-\frac{13}6 + \ln(2) + 2 \gamma\Bigr] z^2 +
\frac{19}{60} \ln(z^2 + 1) + O(z^0) \Biggr\} . \qquad \label{largevector}
\end{eqnarray}
As before, the leading large $z$ form agrees with the flat space result
derived by Coleman and Weinberg \cite{Coleman:1973jx}. Substituting 
(\ref{smallpsi}) in (\ref{vectorDV}) and performing the integral gives
the small field expansion,
\begin{eqnarray}
\lefteqn{\Delta V_{\rm vector} = \frac{3 H^4}{8 \pi^2} \Biggl\{ \frac12 
\ln(1 - \Delta z) + \frac12 \Delta z + \frac14 {\Delta z}^2 + \frac7{12}
{\Delta z}^3 - \frac38 {\Delta z}^4 } \nonumber \\
& & + \sum_{m=1}^{\infty} \zeta(2m+1) \Biggl[-\frac{\Delta z^{2m+1}}{2m+1}
+ \frac{\frac32 {\Delta z}^{2m+2}}{2m+2} + \frac{\frac32 {\Delta z}^{2m+3}}{
2m+3} - \frac{\Delta z^{2m+4}}{2m+4} \Biggr] \Biggr\} , \qquad 
\label{smallvector1}
\end{eqnarray}
where we define $\Delta z$ as,
\begin{equation}
\Delta z \equiv \frac12 - \frac12 \sqrt{1 - 8 z^2} = 2 z^2 + 4 z^4 + 16 z^6
+ 80 z^8 + O(z^{10}) \; . \label{Deltaz}
\end{equation}
Substituting (\ref{Deltaz}) into (\ref{smallvector1}) gives,
\begin{equation}
\Delta V_{\rm vector} = \frac{3 H^4}{8 \pi^2} \Biggl\{
\Bigl(\frac{10}{3} \!-\! \frac83 \zeta(3)\Bigr) z^6 + \Bigl(12 \!-\!
10 \zeta(3)\Bigr) z^8 + O(z^{10}) \Biggr\} \; . \label{smallvector2}
\end{equation}

\section{What Is $H$ Generally?}

In section 2 we saw that various matter couplings to the inflaton induce 
corrections to the inflaton potential which on de Sitter take the general 
form,
\begin{equation}
\Delta V \Bigl\vert_{\rm de\ Sitter} = H^4 f(z^2) \qquad , \qquad
z^2 \propto \frac1{H^2} \; ,
\end{equation}
where $z^2$ is the norm-squared of the coupling constant times the 
inflaton divided by $H$. If inflation were exactly de Sitter then 
$H(\eta)$ would be a constant and we could eliminate all or part of
$\Delta V$ by fine tuning the classical potential. However, $H(\eta)$ 
must change in any realistic model of inflation, and this calls into 
question what functional of the metric those factors of ``$H$'' 
really are, at least for a general inflationary background
of the form (\ref{cosmogeom}). We cannot answer this question by
direct computation because the scalar propagator $i\Delta[\xi,M^2](x;x')$
is not known for geometries other than de Sitter (and flat space)
when $M^2 \neq 0$. We shall instead use indirect arguments to conclude
first, that most of the factors of $H$ are dynamical functionals of the
metric and second, that these functionals cannot even be local.

First, consider the contribution to the stress tensor from whatever 
coupling (scalar, fermion or vector) we choose, 
\begin{equation}
\Delta T_{\mu\nu}(x) \equiv -\frac{2}{\sqrt{-g(x)}} 
\frac{\delta S_{\rm coupling}}{\delta g^{\mu\nu}(x)} \; .
\end{equation}
We can evaluate the expectation value of $\Delta T_{\mu\nu}$ for the
de Sitter geometry, in the appropriate matter vacuum, and use this to 
probe how the factors of $H$ depend upon the metric. For example, if 
all the factors of $H$ are constant then we would find,
\begin{equation}
H^2 = \frac13 \Lambda \qquad \Longrightarrow \qquad \Bigl\langle 
\Delta T_{\mu\nu} \Bigr\rangle_{\rm de\ Sitter} = -g_{\mu\nu} H^4 
\times \frac12 f(z^2) \; . \label{poss1}
\end{equation}
If the factors of $H$ derive from the Ricci scalar, we would find,
\begin{equation}
H^2 = \frac1{12} R \qquad \Longrightarrow \qquad \Bigl\langle 
\Delta T_{\mu\nu} \Bigr\rangle_{\rm de\ Sitter} = -g_{\mu\nu} 
H^4 \times \frac12 z^2 f'(z^2) \; . \label{poss2}
\end{equation}
The presence in this expression of $f'(z^2)$, rather than $f(z^2)$,
originates from the metric variation of $R$.

The actual result is a combination of both possibilities (\ref{poss1}) 
and (\ref{poss2}) \cite{Janssen:2009pb,Miao:2006pn,Prokopec:2007ak}, 
although the predominant behavior is (\ref{poss2}). Only the factors 
of $H^{D-4}$ which were introduced in the renormalization counterterms 
(\ref{dxiscalar}-\ref{dlambdascalar}), 
(\ref{dxifermion}-\ref{dlambdafermion}) and 
(\ref{dxivector}-\ref{dlambdavector}) are consistent with being 
constants. Changing them from $H$ to an arbitrary renormalization 
scale $\mu$ effects the following changes in the de Sitter effective 
potentials,
\begin{eqnarray}
\Delta V_{\rm scalar} & \longrightarrow & \Delta V_{\rm scalar} +
\frac{H^4}{64 \pi^2} \Bigl[ \frac14 z^4 \!+\! 2 \Delta \xi z^2\Bigr]
\ln\Bigl( \frac{H^2}{\mu^2}\Bigr) \; , \qquad \\
\Delta V_{\rm fermion} & \longrightarrow & \Delta V_{\rm fermion} -
\frac{H^4}{8 \pi^2} \Bigl[ \frac12 z^4 \!+\! z^2\Bigr]
\ln\Bigl( \frac{H^2}{\mu^2}\Bigr) \; , \qquad \\
\Delta V_{\rm vector} & \longrightarrow & \Delta V_{\rm vector} +
\frac{3 H^4}{8 \pi^2} \Bigl[ \frac12 z^4 \!+\! z^2\Bigr]
\ln\Bigl( \frac{H^2}{\mu^2}\Bigr) \; . \qquad
\end{eqnarray}
From expressions (\ref{largescalar}), (\ref{largefermion}) and 
(\ref{largevector}) we see that these replacements remove the
logarithmic dependence upon $H^2$ from the $z^4 \ln(z^2)$ and
$z^2 \ln(z^2)$ terms of the large field expansions, leaving only
the $\ln(z^2)$ term.\footnote{Recall that $z^2$ contains the {\it
inverse} of $H^2$, so the factors of $\ln(z^2)$ in expressions
(\ref{largescalar}), (\ref{largefermion}) and 
(\ref{largevector}) go like $-\ln(H^2) = -\ln(\mu^2) - 
\ln(H^2/\mu^2)$.}

The remaining factors of $H^2$ are unknown functionals of the 
metric which have the property that their first variations are 
indistinguishable from that of $\frac1{12} R$ for de Sitter.
This leaves a vast number of possibilities. One can think of
doing a sort of functional Taylor expansion about de Sitter,
and of course fixing the 0th and 1st order terms in no way 
constrains the remaining ones.  

It is easy to see that the remaining factors of $H^2$ are not local
functionals of the metric. Note first that it is the small $\varphi$ 
expansion which matters for inflation.\footnote{If $\varphi$ had a 
large field minimum we would simply have subtracted this off in the
couplings (\ref{scalarL}), (\ref{fermionL}) and (\ref{vectorL}).}
Hence we can expand the propagator in powers of $\varphi$, which
is also the mass of our general scalar propagator (\ref{genprop}),
\begin{eqnarray}
\lefteqn{ i\Delta[\xi,M^2](x;x) = i\Delta[\xi,0](x;x) 
-i M^2 \!\! \int \!\! d^Dw \sqrt{-g(w)} \, 
\Bigl( i\Delta[\xi,0](x;w) \Bigr)^2 } \nonumber \\
& & \hspace{0cm} + \sum_{n=2}^{\infty} (-i M^2)^n \!\! \int\!\! 
d^Dw_1 \sqrt{-g(w_1)} \; i\Delta[\xi,0](x;w_1) \!\! \int\!\! d^Dw_2 
\sqrt{-g(w_2)} \nonumber \\
& & \hspace{1cm} \times i\Delta[\xi,0](w_1;w_2) \dots \!\! \int\!\! 
d^Dw_n \; i\Delta[\xi,0](w_{n-1};w_n) \; i\Delta[\xi,0](w_n;x) \; . 
\qquad \label{propexp}
\end{eqnarray}
Although we cannot solve for $i\Delta[\xi,0](x;x')$ for a general
cosmological geometry (\ref{cosmogeom}), it is known for the infinite
sub-class of geometries with constant $\epsilon(\eta)$ 
\cite{Janssen:2008px}. Except for $\epsilon(\eta) = 0$ (de Sitter),
these geometries interpolate between different values of $H(\eta)$.
With expression (\ref{propexp}) this is already enough to see that
the factors of ``$H^2$'' are not even local.

\section{Conclusions}

The increasingly tight upper bounds on the tensor-to-scalar ratio have
heightened the unease that many feel over the degree of fine tuning which 
is required to make scalar potential models (\ref{singlescalar}) conform
to the data \cite{Ijjas:2013vea,Guth:2013sya,Linde:2014nna,Ijjas:2014nta}.
But it has hitherto been believed that fine tuning would at least suffice,
however distasteful it might be. We have argued here that there may be a 
fundamental obstacle associated with quantum corrections to the inflaton 
potential from the matter couplings which are needed for reheating. 
Contrary to the simple $\varphi^4 \ln(\varphi^2)$ effective potentials
which are induced in flat space \cite{Coleman:1973jx}, the de Sitter 
space results for loops of other scalars (\ref{scalarDV}), massless
fermions (\ref{fermionDV}) and vectors (\ref{vectorDV}) all exhibit a 
complicated dependence on the dimensionless ratio of the inflaton to the
inflationary Hubble parameter. By considering the expectation value of
the stress tensor specialized to de Sitter we showed that most of these
factors cannot be actual constants but must rather be dynamical functionals
of the metric which change with time as any realistic model of inflation 
evolves. It is only the leading term of the large field expansions 
(\ref{largescalar}), (\ref{largefermion}) and (\ref{largevector}), which
recovers the $H$-independent Coleman-Weinberg form. If the factors of 
$H^2$ which appear in the rest of the effective potential were local 
--- for example, $H^2 \rightarrow \frac1{12} R$, which is consistent with 
the de Sitter stress tensor results \cite{Janssen:2009pb,Miao:2006pn,
Prokopec:2007ak} --- then any undesirable effects could still be fine
tuned away. However, we argued in section 3 that the dependence on the
metric cannot be local, even when specialized to a cosmological 
background (\ref{cosmogeom}). {\it This means that no local counterterm
can fully eliminate quantum corrections to the inflaton potential during
actual inflation.} 

More study is needed to quantify how bad the problem actually is.
For example, if undesirable terms in $\Delta V$ are eliminated at one
instant in time using a local counterterm --- either with the
replacement $H^2 \rightarrow \frac13 \Lambda$ or $H^2 \rightarrow
\frac1{12} R$ --- how large does the residual effect at other times 
become during the $\sim 50$ e-foldings of primordial inflation which 
are required to solve the horizon problem? Although the problem cannot
be solved exactly because we only know the key propagator
$i\Delta[\xi,M^2](x;x')$ for de Sitter, it might be reduced to a 
series of conformal time integrations using the expansion (\ref{propexp}) 
for the class of constant $\epsilon(\eta)$ geometries.\footnote{The 
Schwinger-Keldysh formalism \cite{Schwinger:1960qe,Mahanthappa:1962ex,
Bakshi:1962dv,Bakshi:1963bn,Keldysh:1964ud,Chou:1984es,Jordan:1986ug,
Calzetta:1986ey,Calzetta:1986cq,Weinberg:2005vy,Weinberg:2006ac} would 
be necessary to provide a real and causal form for (\ref{propexp}).}
Because the matter theories (\ref{scalarL}), (\ref{fermionL}) and
(\ref{vectorL}) are all renormalizable, the terms at order $M^6$ and 
higher must be ultraviolet finite, so they can be evaluated with 
$D=4$, and the resulting temporal integrations could at least be
performed numerically for specific values of $\epsilon$. 

We expect that the constant $\epsilon$ study would reveal deficiencies 
with the replacement $H^2 \rightarrow \frac1{12} R$ which could be 
repaired with a better local replacement of the form $H^2 \rightarrow
\frac1{12} f(\epsilon) R$, where ``$\epsilon$'' can be reconstructed 
from ratios of curvatures using the constant $\epsilon$ relations,
\begin{eqnarray}
R \longrightarrow 6 (2 \!-\! \epsilon) H^2  & , &
R^{\mu\nu} R_{\mu\nu} \longrightarrow 12 (3 \!-\! 3 \epsilon
+ \epsilon^2) H^4 \; , \qquad \\
\square R \longrightarrow 36 \epsilon (1 \!-\! \epsilon)
(2 \!-\! \epsilon) H^4 & , &
R^{\mu\nu\rho\sigma} R_{\mu\nu\rho\sigma} \longrightarrow
12 (2 \!-\! 2 \epsilon \!+\! \epsilon^2) H^4 \; . \qquad
\end{eqnarray}
Had we not been assuming $\dot{\epsilon} = 0$ the expression for 
$\square R$ would have involved the second slow roll parameter
$\eta = \epsilon - \dot{\epsilon}/2 H \epsilon$, which can be much
larger than $\epsilon$. It is therefore safer to infer ``$\epsilon$'' 
from the other curvature scalars, for example,
\begin{equation}
\frac2{2 \!-\! \epsilon} \longrightarrow 1 + \sqrt{1 \!-\! 
\frac{6 \mathcal{G}}{R^2} } \; ,
\end{equation}
where $\mathcal{G} \equiv R^{\mu\nu\rho\sigma} R_{\mu\nu\rho\sigma}
- 4 R^{\mu\nu} R_{\mu\nu} + R^2$ is the Gauss-Bonnet scalar. Of course
higher curvatures beyond $F(R)$ models would suffer the Ostrogradskian
instability and are not permitted in fundamental theory 
\cite{Woodard:2006nt}.

The data on $n_s$ and $r$ are now inconsistent with any constant 
$\epsilon$ model so even the unacceptable higher curvature terms
will not suffice. However, we should quantify the deviations between 
the best local constant $\epsilon$ subtraction and the actual result 
of $i\Delta[\xi,M^2](x;x)$ for nonconstant $\epsilon$. The tree order 
power spectrum contains a truly nonlocal ``memory effect'' which is 
not recovered by the constant $\epsilon$ result \cite{Wang:1997cw}, 
and the same must be true for $i\Delta[\xi,M^2](x;x)$ as well.

Recent progress on expressing the mode functions for general 
$\epsilon(t)$ holds out the prospect of being able to develop a
comparably effective approximation of the relevant propagators
\cite{Brooker:2015iya}. It is also worthwhile examining the
possibility of synergies between the infrared results we have for
de Sitter and the local expansions which have recently been applied
to Higgs inflation \cite{Barvinsky:2008ia,Barvinsky:2009ii,
Bezrukov:2010jz}. It may be that combining the two methods will 
reveal more about what ``$H^2$'' is generally than either does by 
itself.

A final observation is that these matter loop corrections to the
inflaton effective potential might also modify gravity at late
times because the metric dependence in ``$H^2$'' is not 
Planck-suppressed. We need to quantify how large the residual 
effect can be after the best allowable counterterm has been removed.
In particular, one needs to understand whether or not the nonlocality 
of the primitive contributions can result in some very high scale 
from early times surviving to affect late time physics. 

\noindent {\bf Acknowledgements}

We have benefited from conversations with R. H. Brandenberger, 
T. Prokopec and N. C. Tsamis. This work was partially supported by 
Taiwan MOST grant 103-2112-M-006-001-MY3, by NSF grants PHY-1205591 
and PHY-1506513, and by the UF's Institute for Fundamental Theory.

\end{document}